\documentclass[11pt]{article}

\usepackage[preprint]{acl}

\usepackage{times}
\usepackage{latexsym}
\usepackage{amsmath}

\usepackage[T1]{fontenc}
\usepackage[utf8]{inputenc}

\usepackage{microtype}

\usepackage{inconsolata}

\usepackage[greek,english]{babel}
\usepackage{textgreek}
\usepackage[LGR,T1]{fontenc}
\DeclareFontSubstitution{LGR}{cmr}{m}{n}
\usepackage{graphicx}
\usepackage{multirow}
\usepackage{booktabs}
\usepackage{longtable}   % allows multipage tables
\usepackage{tabularx}
\usepackage{float}
\usepackage[most]{tcolorbox}
\usepackage[dvipsnames]{xcolor}
\usepackage[table]{xcolor}
\usepackage{soul}

\usepackage{minted}
\setminted{
  frame=single,
  bgcolor=gray!10,
  fontsize=\footnotesize,
  autogobble,
  breaklines=true,
  breakanywhere=true,
  breaksymbolleft={},
  breaksymbolright={},
  breaksymbolindentleft=0pt,
  breaksymbolindentright=0pt,
  breaksymbolsepleft=0pt,
  breaksymbolsepright=0pt
}
\newcommand{\framework}{\textsl{EARS}}

\newcommand{\best}[1]{\textbf{#1}}
\newcommand{\secondbest}[1]{\underline{#1}}
\usepackage{wrapfig}

\title{\framework{}: Explanatory Abstention for Reliable Sub-Agent Modeling in Large-Scale Multi-Agent Systems}

\author{
 \textbf{Shuang Xie \thanks{~~denotes equal contribution.}\textsuperscript{1}},
 \textbf{Yunan Lu $^{*}$\thanks{~~This work was done as part of internship at Shopify.}\textsuperscript{1,2}},
 \textbf{Han Li $^{*}$\textsuperscript{1}},
 \textbf{Lingyun Wang \textsuperscript{1}}
\\
 \textsuperscript{1}Shopify \space
 \textsuperscript{2}Columbia University
\\
   \textbf{Correspondence:} \href{mailto:email@domain}{lingyun.wang@shopify.com}
}

\begin{document}
\maketitle
\begin{abstract}
In large-scale enterprise settings, centralized multi-agent systems (MAS) are increasingly adopted, in which a coordinator delegates user requests to lightweight, domain-specialized sub-agents. While this architecture improves modularity, scalability, and cost efficiency, its reliability depends not only on accurate routing but also on sub-agents' ability to calibrate their responses to capability constraints. In particular, sub-agents built on smaller fine-tuned models often struggle with such calibration, leading them to over-answer ambiguous, underspecified, misrouted, or unsupported requests and produce hallucinated outputs instead of actionable feedback.
To address this challenge, we present \framework{} (\underline{E}xplanatory \underline{A}bstention for \underline{R}eliable \underline{S}ub-Agent Modeling), a production-oriented framework that reframes sub-agent abstention as an inter-agent communication protocol: a sub-agent does not merely abstain, but exposes an actionable failure state to the coordinator.
\framework{} curates human-agent interaction data using an ensemble of calibrated LLM-as-a-Judge models, producing structured abstention labels and rationales under a taxonomy of sub-agent failure modes. These data are used to fine-tune sub-agents to detect failure conditions and return rationales for coordinator-level clarification, rerouting, or fallback.
We evaluate \framework{} in a large-scale production e-commerce assistant supporting enterprise business intelligence workflows. \framework{} improves the overall response pass rate from 68.5\% to 78.9\%, demonstrating that sub-agent-side explanatory abstention improves MAS reliability.
% LLM-based applications are increasingly deployed as centralized multi-agent systems (MAS), where a coordinator delegates user requests to lightweight, domain-specialized sub-agents. While this architecture improves modularity, scalability, and cost efficiency, reliability depends on accurate routing and sub-agents' ability to recognize and communicate the capability boundaries.
% In production enterprise settings, smaller fine-tuned sub-agents may over-answer ambiguous, underspecified, misrouted, or unsupported requests, producing hallucinated outputs rather than actionable feedback.
\end{abstract}

\section{Introduction}
% \vspace{-0.4em}
\label{sec:introduction}
% 3/4
Multi-Agent Systems (MAS) have demonstrated impressive capabilities across complex tasks, including software engineering, scientific discovery, data analytics, web automation, and general-purpose problem solving \cite{qian_chatdev_2024, wang_openhands_2025, gottweis_towards_2025, swanson_virtual_2024, fourney_magentic-one_2024}. In industry, a widely adopted design is the centralized MAS architecture \cite{zhang_agentorchestra_2026, wu_autogen_2024, fourney_magentic-one_2024}, in which a powerful coordinator interprets user requests, routes them to specialized sub-agents, and integrates their outputs into a final response. To balance task performance and operational cost, sub-agents often use smaller, domain-specific models fine-tuned for particular domains or databases \cite{abdin_phi-4_2024, lepagnol_small_2024}. However, this architecture introduces reliability challenges: the system can fail not only when the coordinator misroutes a request, but also when the sub-agent returns unfaithful execution results.

% Existing work has predominantly addressed MAS reliability at the coordinator layer. Recent studies improve task planning, model routing, and collaboration patterns to help the coordinator assign work more effectively \cite{yue_masrouter_2025, wu_talk_2026, zhang_agentrouter_2025, dang_multi-agent_2025}. While these methods are vital for mitigating routing errors, they focus on top-down coordination, leaving sub-agent execution failures and the resulting communication breakdown between sub-agents and the coordinator largely unexplored.

% Existing work has predominantly addressed MAS reliability at the coordinator layer. RIRS \cite{wu_talk_2026} selects specialist agents based on knowledge boundaries and decomposes complex queries through planning, while MasRouter \cite{yue_masrouter_2025} formulates MAS routing as a unified decision problem over collaboration modes, role allocation, and LLM selection. Recent cost-aware orchestration work further trains centralized controllers to coordinate expert models under performance--cost trade-offs \citep{jin_controlling_2025}.
% While these methods are vital for mitigating routing errors, they focus on top-down coordination, leaving sub-agent execution failures and the resulting communication breakdown between sub-agents and the coordinator largely unexplored.

Existing work has predominantly addressed MAS reliability at the coordinator layer by improving task planning, agent routing, and collaboration strategies \cite{yue_masrouter_2025, wu_talk_2026, zhang_agentrouter_2025, dang_multi-agent_2025, jin_controlling_2025}. Although these methods mitigate routing and coordination errors, they leave a complementary failure mode underexplored: even a correctly selected sub-agent may over-answer when a request is ambiguous or lacks required information. In these cases, an executable-looking output can propagate an error to the coordinator instead of revealing the condition that prevented faithful execution.

Recent analyses of MAS failures suggest that sub-agent execution and communication are central to system reliability. The failure taxonomy of \citet{cemri_why_2025} shows that many system-level failures arise from breakdowns in inter-agent information flow, including failures to seek clarification, information withholding, and ignoring inputs from other agents. These findings highlight that what a sub-agent returns to the coordinator is not merely an intermediate payload; it is a communication signal for downstream recovery. Prior work has studied when models should refuse unanswerable or unsupported questions \citep{kirichenko_abstentionbench_2025,amayuelas_knowledge_2024}. However, existing methods primarily study single-agent systems, where abstention serves as a final user-facing refusal, rather than a communication signal to the  coordinator.

To address this gap, we propose \framework{}, a domain-adaptive approach for improving the robustness of centralized MAS. \framework{} uses calibrated LLM-as-a-Judge models with unanimous judge agreement to curate abstention supervision from human-agent interaction data. It then fine-tunes sub-agents to return category-specific abstention decisions and rationales as actionable coordinator signals, improving response reliability.

We make three contributions. First, we formulate explanatory abstention at the sub-agent level as a coordination signal for production MAS, turning abstention from a user-facing refusal into actionable inter-agent feedback. Second, we introduce a data curation pipeline to iteratively identify interaction states where sub-agents should abstain. 
Third, we evaluate \framework{} on a large-scale e-commerce platform that serves millions of customers globally and supports substantial annual commerce volume.
\framework{} improves the response pass rate by 15.2\% relatively, demonstrating that explanatory abstention provides a practical mechanism for improving reliability in large-scale centralized MAS.
\section{Related Work}
\label{sec:related}
% 1/2

\subsection{Centralized Multi-Agent Systems}

Centralized MAS coordinate multiple specialized agents through a controller that decomposes requests, assigns tasks, and aggregates intermediate outputs. Frameworks such as AutoGen \cite{wu_autogen_2024}, Magentic-One \cite{fourney_magentic-one_2024}, and AgentOrchestra \cite{zhang_agentorchestra_2026} provide general mechanisms for organizing agent roles and communication workflows. Building on these architectures, recent work has improved how coordinators select and manage specialized agents. RIRS selects specialists according to their knowledge boundaries and combines routing with planning for complex cross-domain queries \citep{wu_talk_2026}. MasRouter jointly determines the collaboration mode, agent roles, and underlying LLMs to balance task performance and inference cost \citep{yue_masrouter_2025}. Other approaches learn routing policies or centralized control strategies under performance-cost trade-offs \citep{zhang_agentrouter_2025, dang_multi-agent_2025, jin_controlling_2025}. These methods primarily improve decisions made before or during task delegation. However, accurate routing does not guarantee faithful execution after a task reaches a sub-agent. A selected sub-agent may lack necessary context, tools, schema coverage, or domain capabilities while still producing a plausible output. \framework{} addresses this complementary problem at the sub-agent--coordinator interface. Rather than changing how the coordinator selects a sub-agent, it trains the selected sub-agent to expose its execution boundary through a structured and actionable response. Consequently, \framework{} can complement existing routing and orchestration methods.

\subsection{Abstention in LLMs}

Abstention enables LLMs to avoid unsupported responses when available knowledge or evidence is insufficient.
One line of work evaluates whether LLMs abstain appropriately under knowledge uncertainty, unanswerable questions, incomplete evidence, ambiguity, and unsupported requests \citep{wen_know_2025, brahman_art_2024, amayuelas_knowledge_2024, madhusudhan_llms_2024}. AbstentionBench broadens this evaluation across diverse forms of unanswerability and shows that even strong reasoning models frequently answer when they should abstain \citep{kirichenko_abstentionbench_2025}. RefusalBench constructs diagnostic cases by applying controlled linguistic perturbations to answerable questions, enabling systematic evaluation across different sources and degrees of informational uncertainty \citep{muhamed_refusalbench_2025}. ConfuseBench specifically focuses on cases involving uncertainty and reveals that LLMs struggle to accurately identify its root cause \cite{liu2025not}. These benchmarks provide controlled and reproducible evaluation signals, but they typically define answerability under a fixed task context. Moreover, synthetically perturbed queries may not fully reflect naturally occurring user requests or the system-specific constraints that make them unanswerable in practice.

Beyond evaluation, recent methods also improve abstention through model training or inference-time intervention. Learning to Clarify introduces action-based contrastive self-training to teach LLMs appropriate dialogue policies, particularly when to ask clarifying questions rather than guess the user’s intent \citep{chen_learning_2025}. Abstain-R1 uses reinforcement learning with verifiable rewards to jointly optimize abstention and follow-up clarification for underspecified requests \citep{zhai_abstain-r1_2026}. LatentRefusal formulates unanswerable Text-to-SQL requests as an answerability-gating problem and detects them from intermediate model representations \citep{ren_latentrefusal_2026}. These approaches improve whether and how an individual model declines a request. Even when they generate clarification, however, they generally formulate recovery as an interaction between the model and the end user. \framework{} studies a different system role: the abstaining model is a specialized sub-agent, and its response becomes an internal coordination signal. The signal must therefore identify not only that execution should stop, but also why it should stop and what action the coordinator can take next.

\begin{figure*}[!ht]
    \centering
    \includegraphics[scale=0.50]{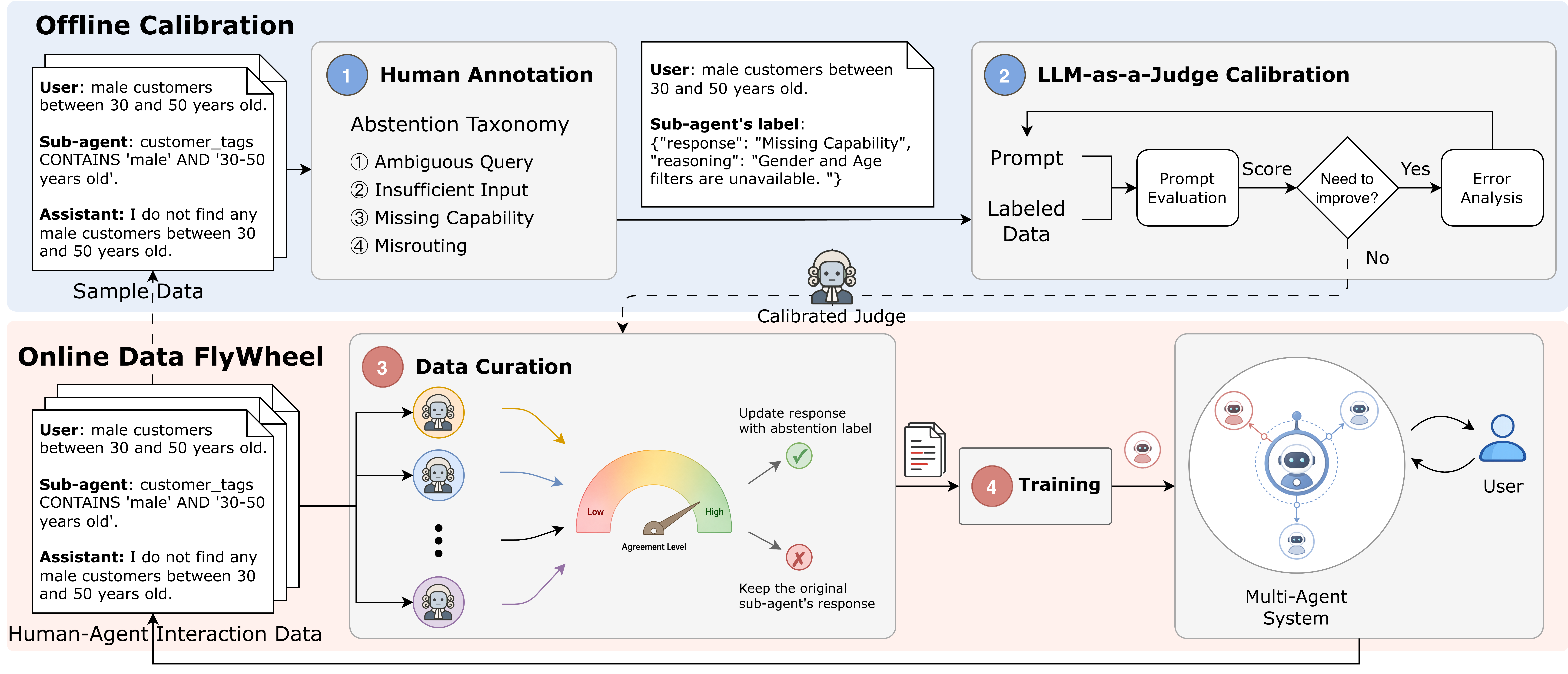}
    % \vspace{-0.5em}
    \caption{Overview of \framework{}. The framework starts by sampling a subset of human-agent interaction data for human annotation, then uses the annotated seed set to calibrate LLM-as-a-Judge models under an abstention taxonomy. The calibrated judges then curate large-scale interaction data for sub-agent fine-tuning. The updated sub-agent is finally redeployed into the multi-agent system, forming an online data flywheel for iterative improvement.}
    \label{fig:pipeline}
    % \vspace{-1.2em}
\end{figure*}

% \vspace{-0.4em}
\section{Method}
% \vspace{-0.4em}
\label{sec:our_method}
% 1.5 pages

We introduce \framework{}, a framework for enhancing MAS resilience by enabling sub-agents to abstain and provide refusal rationales to the coordinator. 
As illustrated in Figure~\ref{fig:pipeline}, \framework{} consists of two stages. The \textbf{offline calibration stage} builds a human-annotated seed dataset with an abstention taxonomy (Section~\ref{sec:human_annotation}), and uses it to calibrate an LLM-as-a-Judge via iterative refinement (Section~\ref{sec:calibration}). 
The \textbf{online data flywheel stage} applies the calibrated judges to curate interaction data (Section~\ref{sec:data_curation}) for sub-agent fine-tuning (Section~\ref{sec:training}). The updated sub-agent is then deployed back into the MAS to collect new interactions and drive the next cycle, enabling iterative improvement.

% \vspace{-0.4em}
\subsection{Human Annotation}

\label{sec:human_annotation}
\paragraph{Abstention Taxonomy.}
To make sub-agent abstention feedback useful for the coordinator's final response, we design a taxonomy that captures why the selected sub-agent cannot handle a delegated request. Building on prior work \cite{wen_know_2025}, we classify abstention cases into four mutually exclusive categories: \emph{Ambiguous Query} \cite{asai_challenges_2021}, where the user intent cannot be clearly determined; \emph{Insufficient Input} \cite{li_mediq_2024}, where the intent is clear but execution requires additional information; \emph{Missing Capability} \cite{amayuelas_knowledge_2024}, where the sub-agent does not support the requested functionality; and \emph{Misrouting}, where the request should be handled by another sub-agent. This taxonomy provides fine-grained failure categories that sub-agents return to the coordinator. We provide examples of each category in Table~\ref{tab:case-study-abstention-examples}.
% \vspace{-0.6em}
\paragraph{Annotation Process.}
To minimize annotation ambiguity and guide annotators through logically dependent label decisions, we structure the annotation process as a hierarchical decision procedure. 
Annotators first determine whether the user query is ambiguous. If the query admits multiple materially different interpretations due to underspecified key information, annotators label it as \emph{Ambiguous Query}. Otherwise, the query is treated as non-ambiguous, meaning that it has a single intended operation.
For non-ambiguous queries, annotators then assess whether the target sub-agent can fulfill the request given its available context, supported capabilities, and business policy constraints. If the sub-agent can complete the request, the instance is marked as solvable. If not, annotators assign it to one of the abstention categories: \emph{Insufficient Input}, \emph{Missing Capability}, or \emph{Misrouting}.

Each annotation object includes both a category label and a written rationale. The rationale specifies the source of ambiguity for \emph{Ambiguous Query}, the missing information for \emph{Insufficient Input}, the unsupported functionality for \emph{Missing Capability}, or the appropriate sub-agent for \emph{Misrouting}. 
% This rationale makes the label auditable and provides direct supervision for training sub-agents.

Following standard annotation practice \cite{lia_read_2025, artstein_survey_2008}, we adopt an independent dual-annotation scheme with adjudication. Each instance is labeled independently by two annotators. Agreed labels are accepted directly; disagreements are resolved by a third senior annotator. Annotation was conducted by a professional data annotation vendor. The process yielded 600 solvable queries and 602 abstention instances, suggesting that a substantial proportion of user queries are not correctly handled by the sub-agent.

% \vspace{-0.5em}
\subsection{LLM-as-a-Judge Calibration}

\label{sec:calibration}

To curate abstention-aware training data from production data at scale, we use an automatic judge to determine whether a query is solvable and, if not, which abstention category it belongs to. Directly applying an off-the-shelf LLM-as-a-judge is unreliable in this setting because abstention decisions depend on task-specific rules and subtle distinctions among abstention categories. We therefore calibrate the judge on the human-annotated seed set before applying it to large-scale data.

Given the human-annotated seed set, we iteratively optimize the LLM-as-a-judge instructions to improve abstention-category prediction accuracy. We begin with an initial general instruction prompt and run the judge on the annotated dataset to measure its performance. If prediction accuracy falls below a predefined threshold, e.g., 80\%, we use a second LLM to analyze the failure cases, summarize prominent confusion patterns, and generate actionable suggestions for revising the guidelines. 
% This establishes a human-in-the-loop revision process that mitigates the risk of overfitting associated with fully automated prompt optimization, while ensuring the guideline updates remain general, interpretable, and consistent with the annotation taxonomy. 
This establishes a human-in-the-loop revision process that mitigates the risk of overfitting associated with fully automated prompt optimization. It also ensures that guideline updates remain general, interpretable, and consistent with the annotation taxonomy.
We repeat this process until the score exceeds the threshold and shows no improvement for two consecutive revisions. 
% \textcolor{red}{\st{To avoid overfitting to the seed set, we do not adopt automatic prompt optimization. Instead, we manually revise the guidelines based on the error analysis, ensuring that each change remains general, interpretable, and consistent with the annotation taxonomy.}}

% \vspace{-0.5em}
\subsection{Data Curation}
\label{sec:data_curation}
We apply the calibrated LLM-as-a-judge to annotate the human-agent interaction data. Prior work shows that different LLMs can exhibit complementary strengths, motivating ensemble-based approaches that combine predictions across models \cite{verga_replacing_2024}. Following this insight, we instantiate judges from three model families: \texttt{OpenAI}~\cite{openai_gpt5_2025}, \texttt{Claude}~\cite{anthropic_claude_sonnet_4_5_2025}, and \texttt{Gemini}~\cite{google_gemini_2_5_flash_2025}. To avoid incorrectly converting answerable queries into abstention cases, we adopt a conservative ensemble strategy. For each sample, we run all judges and accept the label only when they unanimously agree on the same abstention category.

The number of required agreeing judges provides tunable precision--recall control. In our setting, we use unanimous agreement to construct a high-precision curated dataset for training. In applications where recall is more important, such as surfacing potential abstention cases for human review, this threshold can be lowered according to deployment requirements and risk tolerance.

% \vspace{-0.5em}
\subsection{Training}
\label{sec:training}
After curating abstention-aware interaction data, we fine-tune the sub-agent with supervised fine-tuning (SFT). Each training instance consists of the user query, the sub-agent context, and the expected sub-agent response: a task-specific output for solvable queries, or a structured abstention response with an abstention category and a concise rationale. We jointly train on solvable and abstention examples so that the sub-agent learns both to execute supported requests and to recognize its execution boundary, rather than defaulting to either over-answering or excessive abstention.

Subsequently, the sub-agents are deployed back into the MAS to serve live interactions. Newly collected interactions are subsequently returned to the data curation stage, where the calibrated judges identify additional abstention cases for future training cycles, completing the flywheel loop.

\section{Experiments and Results}
\label{sec:experiments}

% \vspace{-0.4em}
\subsection{Experimental Setup}
We evaluate \framework{} on an enterprise-level AI assistant deployed in a production e-commerce environment, focusing on its ability to answer merchant queries related to shop operations and management. The assistant follows a centralized coordinator–sub-agent architecture, where a powerful coordinator built on \texttt{Claude-sonnet-4-6} \cite{anthropic_claude_sonnet_4_6_2026} routes user queries to a set of in-house fine-tuned sub-agents built on \texttt{Qwen3-32B-Instruct} \cite{qwen3-technical-report}. 
% In this work, we focus on the sub-agents in \textit{business intelligence (BI)} domain, which answers data-driven queries about shop performance metrics and operational analysis by generating responses in a Domain-Specific Language (DSL) with SQL-like syntax and functionality. It includes two sub-agents: \textbf{customer segmentation sub-agent} (e.g., How many customers visited my shop in the past week?) and \textbf{business analytics sub-agent} (e.g., What is my shop's conversion rate for the last month?).
In this work, we focus on two sub-agents in the \textit{business intelligence (BI)} domain: the \textbf{customer segmentation sub-agent}, which constructs customer cohorts based on merchant-specified conditions (e.g., Find customers who visited my shop in the past week but did not make a purchase), and the \textbf{business analytics sub-agent}, which answers questions about shop performance and operational metrics by generating responses in a Domain-Specific Language (DSL) with SQL-like syntax (e.g., What was my shop's conversion rate last month?). 

% \vspace{-0.4em}
\paragraph{Dataset.}
The training dataset consists of 13.8K user-agent interactions randomly sampled from the production environment, including 7.0K customer segmentation queries and 6.8K business analytics queries. We then apply the data curation workflow described in Section~\ref{sec:data_curation} to identify unsolvable queries and their corresponding abstention categories: \textit{Ambiguous Query}, \textit{Insufficient Input}, \textit{Missing Capability}, and \textit{Misrouting}. For evaluation, we construct a balanced dataset covering both customer segmentation and business analytics queries. Table~\ref{tab:dataset_statistics} summarizes the statistics of the training and evaluation datasets.
% \begin{table}[t]
% \centering
% \small
% \setlength{\tabcolsep}{3pt}
% \renewcommand{\arraystretch}{1.08}
% \begin{tabular*}{\columnwidth}{@{\extracolsep{\fill}}llrrr@{}}
% \toprule
% \textbf{Split} & \textbf{Data} & \textbf{Std.} & \textbf{Abst.} & \textbf{Total} \\
% \midrule
% Training & Analytics & 6,000 & 0 & 6,000 \\
% Training & Segmentation & 5,367 & 484 & 5,851 \\
% Training & Human Analytics & 800 & 0 & 800 \\
% Training & Human Segmentation & 600 & 602 & 1,202 \\
% \midrule
% Eval & Analytics & 2,000 & 0 & 2,000 \\
% Eval & Segmentation & 1,524 & 476 & 2,000 \\
% \bottomrule
% \end{tabular*}
% \caption{\textcolor{red}{The statistics of training and evaluation dataset. Training rows report the curated training mixture. The human segmentation and analytics rows reports the subset annotated by humans for training. The evaluation set does not provide ground truth standard and abstention labels.(which one is better?)}}

% \begin{table}[H]
% \centering
% \setlength{\tabcolsep}{3pt}
% \renewcommand{\arraystretch}{1.08}
% \begin{tabular*}{\columnwidth}{@{\extracolsep{\fill}}lcc@{}}
% \toprule
% \textbf{Category} &
% \textbf{TrainSet} &
% \textbf{EvalSet} \\
% \midrule
% Solvable & 12,767 &  3,524 \\
% Ambiguous Query &  &  12 \\
% Insufficient Input &  & 103   \\
% Missing Capability &  & 284 \\
% Wrong Sub-agent &  &  77 \\
% \midrule
% \midrule
% Total & 13,853 &  4,000 \\
% \bottomrule
% \end{tabular*}
% \caption{\textcolor{red}{The statistics of training and evaluation dataset. (which one is better?)}}
% \label{tab:statistics}
% \end{table}

\begin{table}[H]
\centering
\setlength{\tabcolsep}{3pt}
\renewcommand{\arraystretch}{0.9}
\setlength{\aboverulesep}{0.3ex}
\setlength{\belowrulesep}{0.3ex}

\newcommand{\subcat}[1]{\hspace{1em}#1}
\newcommand{\grouphead}[1]{\multicolumn{3}{l}{\textbf{\textit{#1}}} \\}

\begin{tabular*}{\columnwidth}{@{\extracolsep{\fill}}lcc@{}}
\toprule
\textbf{Category} & \textbf{Training Set} & \textbf{Eval Set} \\
\midrule
Segmentation Solvable   & 5,967 & 1,524 \\
Analytics Solvable   & 6,800 & 2,000 \\
Abstention & 1,086 & 476 \\
\midrule
\midrule
\textbf{Total} & 13,853 & 4,000 \\
\bottomrule
\end{tabular*}
% \vspace{-0.3em}
\caption{Dataset statistics for training and evaluation.}
% \vspace{-1em}
\label{tab:dataset_statistics}
\end{table}

% \vspace{-0.4em}
\paragraph{Training Setup.}
We fine-tune the sub-agents, built upon \texttt{Qwen3-32B-Instruct}~\cite{qwen3-technical-report}, using supervised fine-tuning (SFT). As the task requires learning nuanced abstention boundaries across multiple categories, we perform full-parameter fine-tuning. We adopt AdamW with a cosine learning rate schedule, setting the peak learning rate to $1\times10^{-5}$ with a warmup ratio of 5\%. We train for three epochs with a global batch size of 8 and a maximum sequence length of 32,768 with data packing. All experiments are conducted on 8 NVIDIA H200 GPUs.

% \vspace{-0.5em}
\subsection{Evaluation Metrics}

\paragraph{Abstention Ability}
Following prior work \cite{feng_dont_2024}, we evaluate the LLM-as-a-Judge's quality in identifying abstention categories using standard metrics: accuracy, precision, and recall.

% \vspace{-0.5em}
\paragraph{Cost} 
We measure computational cost by token usage and normalize it relative to a single \texttt{gemini-2.5-flash} LLM-as-a-Judge call.

% \vspace{-0.5em}
\paragraph{Pass Rate}

We measure sub-agent performance by Pass Rate, defined as the percentage of queries that receive correct responses. The \textbf{Overall Pass Rate} is determined through two sequential verification steps:
(1) \textbf{Syntax Validity (Syntax)}, where a deterministic, code-based verifier checks whether the structured output is valid and executable when applicable, including either a syntactically correct query in the corresponding sub-agent's Domain-Specific Language (DSL) or a properly formatted JSON abstention response; and
(2) \textbf{Semantic Correctness (Semantic)}, where an LLM-as-a-Judge assesses whether the output satisfies the user query along four dimensions: intent alignment, domain constraint satisfaction, query logic correctness, and abstention correctness when applicable.
Outputs that fail the syntax check are automatically marked as overall failures and are not passed to the semantic judge. A response passes overall only if it satisfies both verification steps. We use the same semantic judge across all methods; Appendix~\ref{app:judge-prompts} provides the prompt details.

Because Syntax Validity is determined by a fully reproducible code-based verifier, human agreement is relevant only to the Semantic Correctness judge. To validate this judge, we asked four domain experts with extensive knowledge of the deployed sub-agents to annotate 1,050 production samples using the same rubric. Two experts independently labeled each sample, achieving substantial inter-annotator agreement (Cohen's $\kappa=0.718$), and resolved disagreements through adjudication. Against the resulting consensus labels, the semantic judge achieved 88.3\% accuracy, 82.4\% precision, and 77.8\% recall.

% We measure sub-agent performance by Pass Rate, the percentage of queries that receive correct responses. The \textbf{Overall Pass Rate} is determined by two sequential verification steps:
% (1) \textbf{Syntax Validity (Syntax)}, which checks whether the BI sub-agent's structured output is in a valid format and executable when applicable, including either a syntactically correct SQL-like DSL query or a properly formatted abstention response;
% and (2) \textbf{Semantic Correctness (Semantic)}, which uses an LLM-as-a-Judge to assess whether the output satisfies the user query along four aspects: intent alignment, domain constraint satisfaction, query logic correctness, and abstention correctness when applicable. The same LLM-as-a-Judge is used across all methods. See Appendix \ref{app:judge-prompts} for prompt details.
% A response is considered passing only if both verifications are satisfied. 

% To validate the semantic correctness judge, we asked four domain experts with deep knowledge of the deployed sub-agents to annotate 1,050 production samples using the shared rubrics. Two annotators independently labeled each sample, achieving substantial agreement with Cohen's $\kappa=0.718$, and resolved disagreements through adjudication. Against the consensus labels, the judge achieved 88.3\% accuracy, 82.4\% precision, 77.8\% recall.

\paragraph{Automated Conversation Evaluation (ACE)}

While Pass Rate evaluates sub-agent outputs, we additionally use ACE, our in-house LLM-as-a-Judge system, to assess whether these improvements translate into better final response in the MAS. ACE evaluates each response on a 0-2 scale along five dimensions:
(1) Claim Groundedness, whether domain-specific claims are supported by information available to the model at the corresponding turn;
(2) Goal Fulfillment, whether the system demonstrably completes the user's objective;
(3) User Satisfaction, inferred exclusively from the user's utterances;
(4) Action Execution Success, whether attempted actions complete successfully, independent of whether the system selects the appropriate actions;
(5) Response Usability, whether responses are relevant, comprehensible, and concise.

To validate ACE as an end-to-end evaluator, we compare its scores against a held-out set of 311 conversations independently annotated by domain experts using the same 0--2 rubric. Averaged across the five evaluation dimensions, ACE achieves 82.7\% accuracy, a quadratic weighted Cohen's $\kappa$ of 0.42, and a Kendall's $\tau$ of 0.43. The rank correlations are statistically significant for all dimensions ($p<10^{-3}$), supporting the use of ACE to measure end-to-end system quality. Table~\ref{app:ace-validation} reports the dimension-level results.

% \vspace{-0.4em}
\subsection{Evaluation Results}
% Table~\ref{tab:judge-abstention-ability} reports the calibrated abstention ability of four LLM-as-a-Judge models on 1,202 human-annotated samples. All four judges achieve more than 88\% prediction accuracy. Appendix~\ref{app:judge-calibration} provides detail analysis for each abstention category.

Table~\ref{tab:judge-abstention-ability} reports the calibrated abstention classification performance and relative cost of different LLM-as-a-Judge models on 1,202 human-annotated samples. All four single-model judges achieve above 88\% accuracy, precision, and recall. Among them, \texttt{gemini-2.5-flash} provides the strongest cost-performance trade-off, achieving 90.4\% accuracy at the lowest cost. The 4-judge consensus further improves precision by 6.3\% at higher cost. These results show a practical trade-off between efficiency and accuracy: single-judge curation is preferable when throughput is the primary constraint, while multi-judge consensus provides higher-confidence abstention labels when label quality is the priority. Appendix~\ref{app:judge-calibration} provides a analysis for each abstention category.

% \begin{table}[H]
% \centering
% \setlength{\tabcolsep}{4pt}
% \renewcommand{\arraystretch}{0.95}
% \begin{tabular*}{\columnwidth}{@{\extracolsep{\fill}}lcccc@{}}
% \toprule
% \textbf{LLM-as-a-Judge} &
% \textbf{Accuracy} &
% \textbf{Prec.} &
% \textbf{Recall} &
% \textbf{Cost} \\
% \midrule
% \texttt{gpt-5} & 88.5\% & 89.2\% & 88.5\% & x4.1\\
% \texttt{o3} & 89.6\% & 90.3\% & 89.5\% & x6\\
% \texttt{claude-sonnet-4-5} & 88.4\% & 88.4\% & 88.4\% & x9.3\\
% \texttt{gemini-2.5-flash} & \secondbest{90.4\%} & \secondbest{90.5\%} & \best{90.4\%} & \best{x1} \\
% \midrule
% 4-judge consensus & \best{93.1}\% & \best{96.2}\% & \secondbest{89.8\%} & x20.4\\
% \bottomrule
% \end{tabular*}
% \caption{Abstention classification performance of calibrated LLM-as-a-Judge models. The best result is in \textbf{bold} and the second-best is \underline{underlined}.}
% \label{tab:judge-abstention-ability}
% \end{table}

\begin{table}[t]
\centering
\footnotesize
\setlength{\tabcolsep}{3pt}
\renewcommand{\arraystretch}{0.95}
\begin{tabular}{@{}lcccc@{}}
\toprule
\textbf{LLM-as-a-Judge} &
\textbf{Accuracy} &
\textbf{Precision} &
\textbf{Recall} &
\textbf{Cost} \\
\midrule
\texttt{gpt-5} & 88.5\% & 89.2\% & 88.5\% & x4.1\\
\texttt{o3} & 89.6\% & 90.3\% & 89.5\% & x6\\
\texttt{claude-sonnet-4-5} & 88.4\% & 88.4\% & 88.4\% & x9.3\\
\texttt{gemini-2.5-flash} & \secondbest{90.4\%} & \secondbest{90.5\%} & \best{90.4\%} & \best{x1} \\
\midrule
4-judge consensus & \best{93.1\%} & \best{96.2\%} & \secondbest{89.8\%} & x20.4\\
\bottomrule
\end{tabular}
% \vspace{-0.4em}
\caption{Abstention classification performance of calibrated LLM-as-a-Judge models. The best result is in \textbf{bold} and the second-best is \underline{underlined}.}
% \vspace{-1em}
\label{tab:judge-abstention-ability}
\end{table}

Table~\ref{tab:final-eval-results} reports the response quality of the sub-agents. The baseline corresponds to the MAS with sub-agents trained without abstention data. The prompt-only variant uses the same models and provides the abstention taxonomy in context without additional fine-tuning, whereas \framework{} fine-tunes the sub-agents on data curated by an ensemble of four LLM-as-a-Judge models: \texttt{gpt-5}, \texttt{o3}, \texttt{claude-sonnet-4-5}, and \texttt{gemini-2.5-flash}. Prompt-only abstention increases the Overall Pass Rate from 68.5\% to 70.5\%, while \framework{} further improves it to 78.9\%, outperforming the prompt-only variant by 8.4 percentage points (11.9\% relatively). The gains primarily arise from customer segmentation sub-agent, for which semantic correctness increases from 61.9\% to 82.7\% relative to the baseline. This asymmetry is expected because 23.8\% of queries from customer segmentation sub-agent are predicted to require abstention, compared with fewer than 2\% of analytics queries. Syntax validity remains high across both sub-agents and all variants, suggesting that \framework{} improves abstention decisions without degrading DSL query generation.
% Table~\ref{tab:final-eval-results} reports the response quality of the sub-agents. The baseline corresponds to the MAS with sub-agents trained without abstention data, whereas \framework{} trains the sub-agents on data curated by an ensemble of four LLM-as-a-Judge models built on \texttt{gpt-5}, \texttt{o3}, \texttt{claude-sonnet-4-5}, and \texttt{gemini-2.5-flash}.
% Overall, \framework{} improves most metrics, with larger gains on customer segmentation queries than on analytics queries. This asymmetry is expected since 23.8\% of segmentation queries are predicted as abstention cases, compared with fewer than 2\% of analytics queries. Thus, the segmentation pathway benefits more from abstention training. Syntax validity remains high for both query types across both systems, suggesting that \framework{} improves abstention decisions without degrading DSL query generation quality.
Besides, \framework{} improves the overall ACE score by 2.97\%, indicating that explanatory abstention improves not only sub-agent correctness but also final system behavior.

% \begin{table}[!t]
% \centering
% \small
% \setlength{\tabcolsep}{3pt}
% \renewcommand{\arraystretch}{1.08}
% \begin{tabular*}{\columnwidth}{@{\extracolsep{\fill}}lccc@{}}
% \toprule
% \textbf{Metric} &
% \textbf{Baseline} &
% \textbf{\framework{}} &
% \textbf{Rel. Gain} \\
% \midrule
% Overall Pass Rate & 68.5\% & \best{78.9\%} & \textcolor{Green}{15.2\%} \\
% Seg. Syntax & 92.7\% & \best{97.2\%} & \textcolor{Green}{4.9\%} \\
% Seg. Semantic & 61.9\% & \best{82.7\%} & \textcolor{Green}{33.6\%} \\
% Analytics Syntax & 96.3\% & \best{96.9\%} & \textcolor{Green}{0.6\%} \\
% Analytics Semantic & \best{75.1\%} & 75.0\% & \textcolor{Red}{-0.1\%} \\
% \bottomrule
% \end{tabular*}
% % \vspace{-0.4em}
% \caption{Evaluation results for the BI sub-agent. Seg. and Analytics denote customer segmentation and business analytics queries. Syntax and Semantic denote syntax validity and semantic correctness pass rates. Rel. Gain reports the relative improvement of \framework{} over the baseline, with \textcolor{Green}{green} / \textcolor{Red}{red} indicating gains / losses.}
% % \vspace{-1.5em}
% \label{tab:final-eval-results}
% \end{table}

\begin{table}[!t]
\centering
\small
\setlength{\tabcolsep}{3pt}
\renewcommand{\arraystretch}{1.08}
\begin{tabular*}{\columnwidth}{@{\extracolsep{\fill}}lccc@{}}
\toprule
\textbf{Metric} &
\textbf{Baseline} &
\textbf{Prompt-Only} &
\textbf{\framework{}} \\
\midrule
Overall Pass Rate    & 68.5\% & 70.5\% & \best{78.9\%} \\
Seg. Syntax          & 92.7\% & 93.5\% & \best{97.2\%} \\
Seg. Semantic        & 61.9\% & 65.8\% & \best{82.7\%} \\
Analytics Syntax     & 96.3\% & 96.2\% & \best{96.9\%} \\
Analytics Semantic   & \best{75.1\%} & \best{75.1\%} & 75.0\% \\
\bottomrule
\end{tabular*}
% \vspace{-0.4em}
\caption{Evaluation results for the customer segmentation and business analytics sub-agents. Seg. and Analytics denote customer segmentation and business analytics, respectively. Syntax and Semantic denote syntax validity and semantic correctness pass rates. Prompt-Only provides the abstention taxonomy in context without abstention-aware fine-tuning.}
% \vspace{-1.5em}
\label{tab:final-eval-results}
\end{table}

% \paragraph{Human Evaluation.}
% We further validate abstention behavior with domain-expert annotations at both the sub-agent and full MAS levels. For both evaluations, we use the same targeted abstention set of 584 examples from production traffic, in which experts label whether each request should be handled by abstention. Since all model outputs in this set contain abstention behavior, the binary evaluation measures whether abstention is appropriate rather than exact abstention-reason agreement. On 584 labeled examples, the sub-agent reaches 94.0\% accuracy for abstention detection. Appendix~\ref{app:human-eval-details} reports abstention-category agreement.

% For the end-to-end MAS, experts rate final assistant response quality on the same targeted abstention set with a three-point rubric. We count ratings 1 and 2 as passing responses and rating 0 as failing. On the 584 rated responses, 392 pass, giving a human-rated response-quality accuracy of 67.1\% on the abstention cases. The response-quality distribution is 322 excellent, 70 satisfactory, and 192 poor. Appendix~\ref{app:human-eval-details} reports the response-quality rubric and detailed breakdown.

\begin{figure*}[!ht]
    \centering
    \includegraphics[scale=0.49]{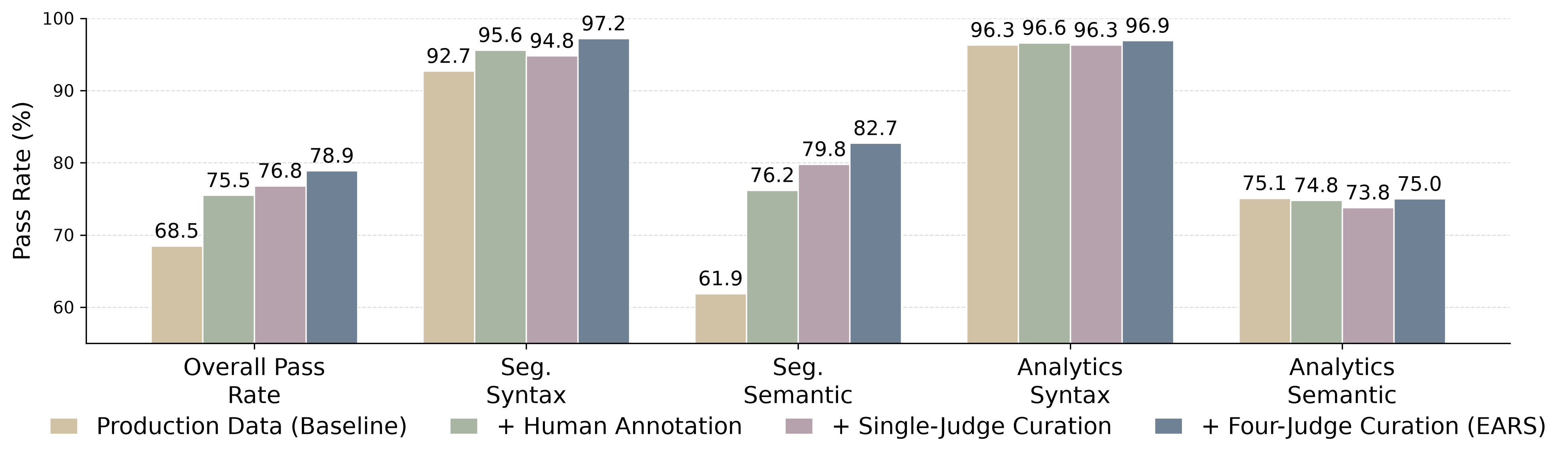}
    % \vspace{-1em}
    \caption{Ablation results on the effect of abstention data under different curation strategies. Adding human-annotated abstention data improves the overall pass rate and substantially increases segmentation semantic correctness. \framework{} uses four-judge curation, which achieves the strongest overall performance.}
    % \vspace{-0.4em}
    \label{fig:ablation_sub_agent_results}
\end{figure*}

% \vspace{-0.5em}
\paragraph{Human Evaluation.}
% To support side-by-side comparison with the incumbent system, we conducted an end-to-end shadow deployment with human validation. From randomly sampled production sessions, we retained 584 examples for which the shadow EARS deployment produced an abstention signal. Independent domain experts then evaluated these examples at two granularities. 

To support side-by-side comparison with the incumbent system, we conducted an end-to-end shadow deployment with human validation. We randomly sampled 2,000 queries from all production sessions routed to either BI-domain sub-agent in June 2026. Third-party domain experts evaluated each example at two granularities. Each example received two independent labels at each granularity, with disagreements adjudicated by a third expert.

At the end-to-end MAS level, experts assessed session-level response quality using a binary rubric, with 1 denoting success and 0 denoting failure. \framework{} achieved a human-rated success rate of 84\%, substantially outperforming the incumbent system's 71.5\%. At the sub-agent level, its abstention decisions achieved 93\% accuracy, 93\% precision, 70\% recall, and 80\% F1, demonstrating strong alignment with expert judgments. Appendix~\ref{app:human-eval-details} provides the complete evaluation rubrics.

% \vspace{-0.5em}
\subsection{Ablations}
To assess the effect of abstention data and LLM-as-a-Judge ensemble strategy, we run ablations across dataset variants with different curation settings. The results are summarized in Figure~\ref{fig:ablation_sub_agent_results}.
Overall, abstention data consistently improves segmentation performance without degrading analytics performance. 
Human-annotated data increases the overall pass rate by 10.2\% relatively. The gain is most pronounced in segmentation semantic correctness, with a 23.1\% relative improvement, demonstrating that abstention examples can improve model reliability.

Although human annotation provides high-quality abstention supervision, mixing it with raw production interactions introduces label inconsistencies. Semantically similar queries may receive conflicting supervision: one is labeled with an abstention category, while another retains the original incorrect sub-agent response. \framework{} curation reduces this conflict by correcting production examples, yielding further gains.

Among single-judge variants, the best-performing judge (\texttt{gemini-2.5-flash}) improves the overall pass rate by 1.3\% over human annotation alone. In contrast, four-judge consensus achieves a 4.5\% gain and improves all four metrics. This pattern suggests that single-judge curation can introduce label noise, whereas consensus-based curation provides a more reliable training signal by retaining only high-agreement corrections.

\subsection{Cost analysis}
We further compare the deployment costs of \framework{} and inference-time gating using the calibrated judges. For 10 million queries, inference-time gating incurs a recurring monthly cost of approximately \$90K for a single \texttt{gemini-2.5-flash} judge at \$0.009 per query, or \$1.83M for the four-judge consensus configuration. In contrast, \framework{} incurs a one-time offline cost of \$3,293, comprising \$2,525 for curating 13.8K training examples and \$768 for 24 hours of training on eight H200 GPUs. The one-time cost of \framework{} is equivalent to processing approximately 366K queries with the single-judge gate and only 18K queries with the four-judge gate. Therefore, although inference-time gating may remain economical at low query volumes, offline fine-tuning substantially reduces costs for a production system serving millions of queries each month.
% \section{Results}
% \label{sec:results}
% 1.5 page

\begin{table*}[!t]
\centering
\small
\setlength{\tabcolsep}{3pt}
\renewcommand{\arraystretch}{1.08}
\newcommand{\diffcase}[1]{\textcolor{BrickRed}{\textbf{#1}}}
\newcommand{\good}[1]{\textcolor{Green}{\textbf{#1}}}

\begin{tabularx}{\textwidth}{
  >{\raggedright\arraybackslash}p{0.09\textwidth}
  >{\raggedright\arraybackslash}p{0.14\textwidth}
  >{\raggedright\arraybackslash}X
  >{\raggedright\arraybackslash}X
}
\toprule
\textbf{Type} &
\textbf{User request} &
\textbf{Baseline} &
\textbf{EARS} \\

\midrule

\multirow[t]{2}{0.07\textwidth}{\emph{Ambiguous Query}} &
\multirow[t]{2}{0.14\textwidth}{"find customers who might be interested in buying shoes wholesale"} &
\textbf{Sub-agent:}
\diffcase{\texttt{storefront.product\_viewed}}\texttt{ MATCHES} \texttt{(tag contains '}\diffcase{\texttt{wholesale}}\texttt{')} &
\textbf{Sub-agent:}
\{\texttt{"Response":\good{"Ambiguous Query"}},
\texttt{"Reasoning": "What is wholesale-interest signal?"}\} \\

& &
\textbf{Assistant:}
"\diffcase{I do not find any customers} who might be interested in buying shoes wholesale..." \newline
\textbf{\underline{Issue}:} Uses an unverified proxy for wholesale buyer interest instead of asking for clarification.
&
\textbf{Assistant:}
"\good{The segmentation tool needs a clearer signal for wholesale interest.} Could you specify what indicates potential wholesale buyers, such as repeat shoe purchases, large orders, or views of wholesale products?" \\

\midrule

\multirow[t]{2}{0.07\textwidth}{\emph{Insufficient Input}} &
\multirow[t]{2}{0.14\textwidth}{"customers who purchased products with type French Horns, Trombones, or Trumpets"} &
\textbf{Sub-agent:}
\texttt{product\_purchased MATCHES}\newline
\texttt{(}\diffcase{\texttt{tag contains}}\texttt{ '}\diffcase{\texttt{Frentch Horns}}\texttt{' OR contains} \texttt{'}\diffcase{\texttt{Trombines}}\texttt{' OR '}\diffcase{\texttt{Trumpets}}\texttt{')} &
\textbf{Sub-agent:}
\{\texttt{"Response": \good{"Insufficient Input"}},
\texttt{"Reasoning": "Need intended product."}\} \\

& &
\textbf{Assistant:}
"\diffcase{I do not find any customers} who purchased products with these types." \newline
\textbf{\underline{Issue}:} Attempts segmentation without resolved product inputs, leading to invalid product filters.
&
\textbf{Assistant:}
"Could you clarify which products you mean by French Horns, Trombones, or Trumpets? \good{I need the intended product} before creating the segment." \\

\midrule

\multirow[t]{2}{0.07\textwidth}{\emph{Missing Capability}} &
\multirow[t]{2}{0.14\textwidth}{"male customers between 30 and 50 years old"} &
\textbf{Sub-agent:}
\diffcase{\texttt{customer\_tags}}\texttt{ CONTAINS '}\diffcase{\texttt{male}}\texttt{'} \texttt{AND }\diffcase{\texttt{customer\_tags}}\texttt{ CONTAINS}\newline
\texttt{'}\diffcase{\texttt{30-50 years old}}\texttt{'} &
\textbf{Sub-agent:}
\{\texttt{"Response": \good{"Missing Capability"}},
\texttt{"Reasoning": "Gender and Age filters are unavailable."}\} \\

& &
\textbf{Assistant:}
"\diffcase{I do not find any male customers between 30 and 50 years old.}" \newline
\textbf{\underline{Issue}:} Maps unsupported demographic attributes to tags, falsely implying searchable gender and age fields.
&
\textbf{Assistant:}
"\good{I cannot create a segment directly from gender or age because these filters are unavailable.} I can help segment customers using supported criteria, such as purchase behavior, location, customer tags, or engagement." \\

\midrule

\multirow[t]{2}{0.07\textwidth}{\emph{Misrouting}} &
\multirow[t]{2}{0.14\textwidth}{"create segment for customers who visited the store multiple times but never purchased"} &
\textbf{Sub-agent:}
\diffcase{\texttt{number\_of\_orders = 0}} &
\textbf{Sub-agent:}
\{\texttt{"Response": \good{"Misrouting"}},
\texttt{"Reasoning": "Need workflow sub-agent."}\} \\

& &
\textbf{Assistant:}
"Here are the customer segments for \diffcase{customers who never purchased}: ..." \newline
\textbf{\underline{Issue}:} Misroutes workflow-dependent request to segmentation, dropping repeated-visit condition.
&
\textbf{Assistant:}
"I routed this request to the workflow sub-agent. \good{It will first tag customers who visit multiple times without purchasing}; then I will create a segment from that tag." \\

\bottomrule
\end{tabularx}
% \vspace{-0.6em}
\caption{Representative abstention examples comparing the baseline and \framework{}. The baseline over-answers ambiguous, underspecified, unsupported, or misrouted requests, whereas \framework{} abstains with category-specific rationales that support more accurate assistant responses. \diffcase{Red} marks errors, and \good{green} marks accurate outputs.}
% \vspace{-2.0em}
\label{tab:case-study-abstention-examples}
\end{table*}

% \vspace{-0.5em}
\section{Case Study}
% \vspace{-0.5em}
To better understand how abstention training affects deployed MAS behavior, we conduct a qualitative analysis over representative cases. Table~\ref{tab:case-study-abstention-examples} compares the baseline with \framework{} across four abstention categories. The examples show that the baseline sub-agent often forces underspecified, unsupported, or misrouted requests into executable segmentation predicates. This behavior propagates to the assistant, which returns a hallucinated or misleading response to the user. In contrast, \framework{} produces category-specific abstention rationales that help the assistant choose a better next action.

For example, in the \emph{Missing Capability} case, the baseline treats demographic attributes as ordinary customer tags, even though gender and age filters are unavailable to the customer segmentation sub-agent due to privacy and compliance policies. This response is problematic because it implies access to sensitive information that the system should not expose. \framework{} instead recognizes the capability boundary and guides the assistant to explain the unsupported filters while suggesting feasible alternatives. Similarly, in the \emph{Ambiguous Query} case, the baseline assumes a proxy for wholesale interest without confirming the user's intended definition. \framework{} identifies this ambiguity and asks the assistant to clarify the signal before segmentation. The \emph{Misrouting} case further shows that abstention can preserve user intent across agent boundaries. By identifying the required workflow sub-agent, \framework{} enables the coordinator to construct the necessary intermediate tag before invoking segmentation sub-agent. This illustrates how explanatory abstention supports multi-step recovery rather than merely preventing an incorrect local execution.

These cases show that abstention rationales not only prevent invalid outputs, but also help the coordinator clarify, redirect, or constrain the task before execution.

% \vspace{-0.4em}
\section{Conclusion}
% \vspace{-0.4em}
\label{sec:conclusion}
% We presented \framework{}, a domain-agnostic method for improving centralized MAS reliability by grounding sub-agent communication through explanatory abstention. It teaches sub-agents to recognize failure cases and return category-specific rationales as coordinator feedback. Using calibrated LLM-as-a-Judge curation and supervised fine-tuning, \framework{} improves the overall response success rate from 68.5\% to 78.9\% on production business intelligence tasks. These results show that sub-agent abstention is not merely a refusal mechanism, but an effective coordination signal for building more reliable deployed MAS.

We presented \framework{}, a production-oriented framework for improving centralized MAS reliability through explanatory sub-agent abstention. Rather than treating abstention as a terminal user-facing refusal, \framework{} reframes it as an inter-agent communication protocol, enabling sub-agents to expose actionable failure states to the coordinator. Using calibrated LLM-as-a-Judge curation and supervised fine-tuning, \framework{} improves the overall response pass rate by 15.2\% relatively in a production e-commerce assistant. These results demonstrate that explanatory abstention provides an effective coordination signal for improving MAS reliability.

\section{Limitations}
\label{sec:limitations}
% Although \framework{} improves the overall response success rate from 68.5\% to 78.9\%, it assumes that the product abstention boundary remains reasonably stable during each curation cycle. In practice, tools, schemas, policies, and supported product capabilities change. These changes can create cases where a previously valid answer path becomes unsupported, or where a previously unanswerable request becomes answerable. This does not change the core method, but it means the curation loop should be rerun after substantial product changes. Teams may need to refresh annotation guidelines, audit seed data, recalibrate the LLM-as-a-judge models, and rebuild the fine-tuning set.

% The method is designed to be domain-agnostic, but our empirical experiment only tested on an e-commerce business-intelligence assistant case. Other domains may have different abstention cases, annotation costs, and coordinator recovery behaviors. Future work should apply the \framework{}'s curation and fine-tuning pipeline on MAS deployed in different domain, such as financial, education and healthcare. RL-based coordinator or recovery-policy optimization could also complement \framework{}, but it is orthogonal to our focus on sub-agent abstention supervision.

The proposed method is designed to be domain-adaptive, while our empirical evaluation focuses on an e-commerce business-intelligence assistant. Further validation is needed in other domains, where abstention scenarios and evaluation costs may differ. Future work should apply \framework{}'s curation and fine-tuning pipeline to MAS deployed in domains such as finance, education, and healthcare.

To obtain reliable abstention labels, we adopt a human-in-the-loop workflow for calibrating LLM-as-a-Judge models. While this design improves interpretability and reduces the risk of overfitting to the seed set, it also motivates more automated calibration strategies. Recent advances in automated judge revision and prompt refinement \cite{wang2024promptagent,agrawal2025gepa} should be explored to improve scalability while preserving label quality.

Finally, RL-based coordinator training or recovery-policy optimization could complement \framework{} by improving MAS-level response strategies following sub-agent abstention. We leave such system-level optimization to future work.

% Our seed data is sampled from production traffic, which worked in our deployment because refusal-worthy segmentation cases appeared often enough to cover the main categories. In deployments where such cases are rare, seasonal, or concentrated in long-tail workflows, random sampling may miss important boundary cases. A practical extension is to combine random sampling with stratified sampling by tool route, active learning over uncertain judge predictions, and targeted mining of user-corrected interactions.

% The method is designed to be domain-agnostic, but our empirical experiment only tested on an e-commerce business-intelligence assistant case. Other domains may have different abstention cases, annotation costs, and coordinator recovery behaviors. We plan xxx.

% Inital version human in the loop process. We plan to use automated prompt improvement, GEPA/automated harness (hermes, openclaw etc) to curate the training data ...

% Future work should apply the \framework{}'s curation and fine-tuning pipeline on MAS deployed in different domain, such as financial, education and healthcare. RL-based coordinator or recovery-policy optimization could also complement \framework{}, but it is orthogonal to our focus on sub-agent abstention supervision.

% \section{Ethics Statement}

\section{Acknowledgments}
We thank Prof. Zhou Yu for valuable guidance. We are also grateful to Jason Liu, Nicolas Bertagnolli, and Aly Gemae for their helpful support and feedback throughout this work. We further thank Kaitlyn Chenier and the annotation team for their assistance with data annotation.

\bibliography{custom}

\clearpage
\onecolumn
\appendix

\lstdefinestyle{segmentationprompt}{
  basicstyle=\ttfamily\tiny,
  backgroundcolor=\color{gray!8},
  breaklines=true,
  breakatwhitespace=false,
  columns=fullflexible,
  keepspaces=true,
  frame=single,
  framerule=0.3pt,
  rulecolor=\color{black!25},
  showstringspaces=false,
  tabsize=2,
  xleftmargin=0.5em,
  xrightmargin=0.5em,
  literate=
    {£}{{\pounds}}1
    {—}{{--}}2
    {‘}{{`}}1
    {’}{{'}}1
    {“}{{``}}2
    {”}{{''}}2
    {…}{{...}}3
    {→}{{$\rightarrow$}}1
    {✓}{{[OK]}}4
    {❌}{{[X]}}3
}

\section{Semantic Correctness Judge Prompts}
\label{app:judge-prompts}

We use task-specific semantic correctness prompts for analytics outputs and segmentation outputs. These prompts are used by the final evaluator, \texttt{openai:gpt-5} with \texttt{temperature=0}, and are separate from the abstention-labeling prompts used by the data-curation judges. The evaluator shares the \texttt{gpt-5} base model with one curation judge, but it is run as a separate evaluation pipeline and is not used to label SFT data. Before final evaluation, the prompt is calibrated on a separate 1,050 example human-annotated split. The production prompts include platform-specific schemas, domain terminology, and internal unsupported-request lists. For anonymized disclosure, we report sanitized excerpts that preserve the evaluation protocol while masking platform names, schema fields, and domain-specific capabilities.

\subsection{Analytics Judge Prompt}
\label{app:analytics-judge-prompt}

\begin{lstlisting}[style=segmentationprompt]
Input:
  user_message: the original user request
  context: request metadata, including request time when available
  model_output: the generated [ANALYTICS_QUERY_LANGUAGE] query

Task:
  Determine whether model_output answers the user's request.
  Return only a JSON object with:
    explanation: detailed reasoning about intent alignment and validity
    result: true if the query is correct, false otherwise

Core checks:
  1. Validate that all metrics, dimensions, and models are allowed by
     [ANALYTICS_SCHEMA_RESTRICTIONS].
  2. Check that the query matches the user's intent, including filters,
     aggregation level, requested entity type, and requested time range.
  3. If the user asks for a visualization, require a supported visualization
     directive from [SUPPORTED_VISUALIZATION_TYPES].
  4. For trend requests, require an ordering or grouping that makes the trend interpretable.
  5. For top, highest, maximum, or best requests, require descending ordering
     by an appropriate metric.
  6. For bottom, lowest, minimum, or least requests, require ascending ordering
     by an appropriate metric.
  7. If no date is specified, treat all-time data as the default.
  8. Interpret relative date ranges using the request timestamp in context.
  9. Apply domain-specific semantic rules for reversal metrics, traffic-source
     dimensions, entity-name matching, and exclusion filters, with all
     platform-specific field names masked.
  10. Reject outputs that attempt unsupported reports listed in
      [UNSUPPORTED_REQUEST_TYPES].
\end{lstlisting}

\subsection{Segmentation Judge Prompt}
\label{app:segmentation-judge-prompt}

\begin{lstlisting}[style=segmentationprompt]
Input:
  user_message: the original user request
  context: request metadata, including request time when available
  model_output: a generated [SEGMENTATION_FILTER_LANGUAGE] expression or
    structured abstention signal

Task:
  Evaluate whether model_output correctly handles the user's request.
  Return only a JSON object with:
    explanation: detailed reasoning about the selected case and final verdict
    result: true if the output is correct, false otherwise

Case selection:
  A. Standard query:
     Use this case when model_output does not contain an abstention signal.
     Check whether the filters, operators, values, and date constraints match
     the user's intent under [SUPPORTED_SEGMENTATION_FILTERS].

  B. Abstention query:
     Use this case when model_output contains an abstention signal.
     Independently decide whether the request is:
       solvable: supported with information already provided
       insufficient_input: supported but missing concrete user-provided input
       not_solvable: unsupported or too ambiguous to execute
     If not_solvable, assign exactly one reason:
       missing_capability
       misrouting
       ambiguous_query
     The model output is correct only when its abstention type and reason match
     the independent assessment.

Core checks:
  1. Confirm intent alignment before judging syntax-level details.
  2. Treat segmentation as audience-list construction, not aggregate analytics.
  3. Reject requests that require aggregate reports, ranking, visualization,
     mutation, or unsupported capabilities in [UNSUPPORTED_REQUEST_TYPES].
  4. Use request time from context for relative dates.
  5. Prefer lifecycle/value-group filters for vague high-value or loyalty
     requests when no numeric threshold is provided.
  6. Do not accept invented tags, unsupported attributes, or proxy filters that
     create unsupported behavior.
\end{lstlisting}

\begin{table}[t]
\centering
\footnotesize
\setlength{\tabcolsep}{1pt}
\renewcommand{\arraystretch}{0.95}
\begin{tabular*}{\columnwidth}
{@{\extracolsep{\fill}}lccc@{}}
\toprule
\textbf{Dimension} & \textbf{Accuracy} & \textbf{Quadratic $\kappa$} & \textbf{Kendall's $\tau$} \\
\midrule
Claim Groundedness       & 70.4\% & 0.30 & 0.38 \\
Goal Fulfillment         & 73.0\% & 0.53 & 0.53 \\
User Satisfaction        & 82.0\% & 0.44 & 0.45 \\
Action Execution Success & 89.1\% & 0.24 & 0.20 \\
Response Usability       & 99.0\% & 0.57 & 0.57 \\
\bottomrule
\end{tabular*}
\caption{Agreement between ACE and human expert annotations on 311 held-out conversations.}
\label{app:ace-validation}
\end{table}

\section{Human Evaluation Details}
\label{app:human-eval-details}

% This appendix reports the response-quality rubric used for the full MAS human evaluation in Section~\ref{sec:experiments}. Domain annotators rated session-level assistant responses with the rubric in Table~\ref{tab:e2e-human-rubric}. Rating 1 is counted as passing responses; rating 0 is counted as failing.

This appendix reports the response-quality rubric used for the full MAS human evaluation in Section~\ref{sec:experiments}. Domain annotators assessed session-level assistant responses according to the rubric in Table~\ref{tab:e2e-human-rubric}. A rating of 1 is counted as a pass, whereas a rating of 0 is counted as a failure.

\begin{table}[H]
\centering
\small
\setlength{\tabcolsep}{5pt}
\renewcommand{\arraystretch}{1.12}
\begin{tabularx}{0.98\textwidth}{p{0.08\textwidth}p{0.18\textwidth}X}
\toprule
\textbf{Score} & \textbf{Label} & \textbf{Rubric} \\
\midrule
% 0 & Poor Quality & The response has one or more serious issues: confusing or illogical explanations, excessive verbosity, multiple unnecessary tool calls, criticism of platform features or recommendations for competitors, or inappropriate tone. \\
% 1 & Satisfactory & The response is generally coherent with minor structure or flow issues, reasonably concise with some unnecessary information, uses tools appropriately most of the time, and maintains a professional tone. \\
% 2 & Excellent & The response is logically organized, concise and focused, makes precise tool calls that directly address the merchant's query, and uses appropriate formatting for readability. \\
0 & Poor Quality & The response has one or more serious issues: confusing or illogical explanations, excessive verbosity, unnecessary or incorrect tool calls, criticism of platform features or recommendations for competitors, inappropriate tone, or failure to provide actionable guidance. \\
1 & Satisfactory & The response is coherent, professionally toned, and sufficiently addresses the user's query. It may contain minor structure, flow, or conciseness issues, but uses tools appropriately and provides actionable guidance. \\

\bottomrule
\end{tabularx}
\caption{Human E2E response-quality rubric.}
\label{tab:e2e-human-rubric}
\end{table}

% This appendix reports the response-quality rubric used for the full MAS human evaluation in Section~\ref{sec:experiments}. Domain annotators rated final assistant responses with the rubric in Table~\ref{tab:e2e-human-rubric}. Ratings 1 and 2 are counted as passing responses; rating 0 is counted as failing.

% \begin{table}[H]
% \centering
% \small
% \setlength{\tabcolsep}{5pt}
% \renewcommand{\arraystretch}{1.12}
% \begin{tabularx}{0.98\textwidth}{p{0.08\textwidth}p{0.18\textwidth}X}
% \toprule
% \textbf{Score} & \textbf{Label} & \textbf{Rubric} \\
% \midrule
% 0 & Poor Quality & The response has one or more serious issues: confusing or illogical explanations, excessive verbosity, multiple unnecessary tool calls, criticism of platform features or recommendations for competitors, or inappropriate tone. \\
% 1 & Satisfactory & The response is generally coherent with minor structure or flow issues, reasonably concise with some unnecessary information, uses tools appropriately most of the time, and maintains a professional tone. \\
% 2 & Excellent & The response is logically organized, concise and focused, makes precise tool calls that directly address the merchant's query, and uses appropriate formatting for readability. \\
% \bottomrule
% \end{tabularx}
% \caption{Human E2E response-quality rubric.}
% \label{tab:e2e-human-rubric}
% \end{table}

\section{LLM-as-a-Judge Calibration Analysis}
\label{app:judge-calibration}

% Tables~\ref{tab:multi-judge-calibration} and
% \ref{tab:abstention-judge-correctness} assess the judges used to curate the abstention
% data. These tables are not part of the main model comparison. They estimate the
% potential accuracy of the judge pool used during curation.

% Table~\ref{tab:multi-judge-calibration} calibrates the LLM judges against human
% annotations. All four judges reach at least 88.4\% binary calibration accuracy. The
% judges also maintain high abstention precision. The lowest abstention precision is
% 87.5\%, and the highest is 95.6\%. These results support the use of conservative
% multi-judge curation for abstention data.

\begin{figure*}[t]
\centering
\includegraphics[width=0.83\textwidth]{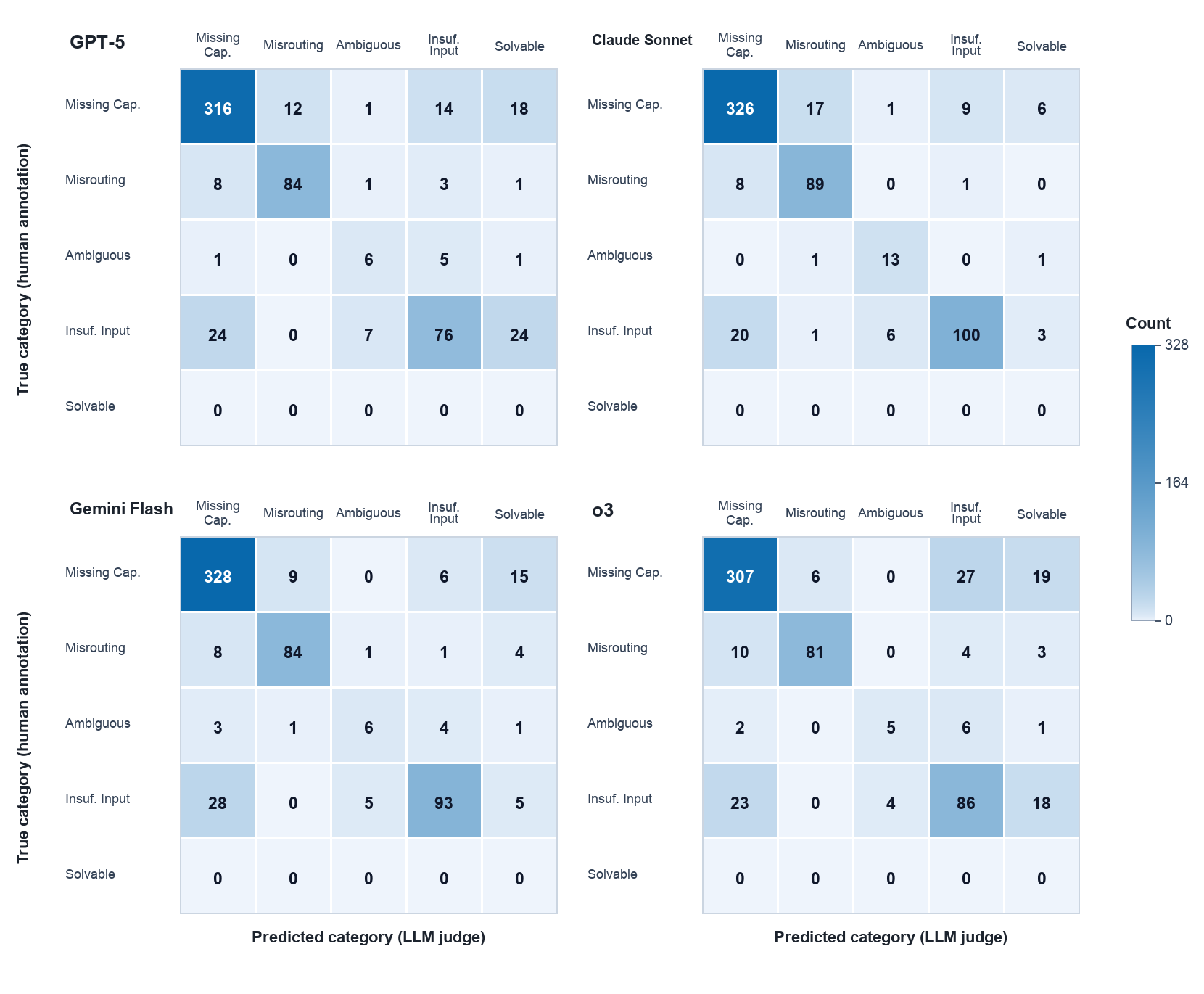}
\vspace{-18pt}
\caption{Abstention judge confusion matrices for final-model abstention-category predictions.
The y-axis gives the true human-annotated category, and the x-axis gives the category
predicted by each LLM judge. 
Darker cells indicate larger counts.}
\label{fig:abstention-judge-confusion-matrices}
\end{figure*}

% \begin{table}[H]
% \centering
% \setlength{\tabcolsep}{4pt}
% \renewcommand{\arraystretch}{1.12}
% \begin{tabular}{lccccc}
% \toprule
% \textbf{LLM Judge} &
% \textbf{Accuracy} &
% \textbf{Std. Prec.} &
% \textbf{Std. Recall} &
% \textbf{Abstain Prec.} &
% \textbf{Abstain Recall} \\
% \midrule
% GPT-5 & 88.5\% & 84.0\% & 95.0\% & 94.3\% & 82.1\% \\
% o3 & 89.6\% & 84.9\% & \best{96.2\%} & \best{95.6\%} & 82.9\% \\
% Claude Sonnet & 88.4\% & \best{89.4\%} & 86.8\% & 87.5\% & \best{90.0\%} \\
% Gemini Flash & \best{90.4\%} & 88.5\% & 92.8\% & 92.4\% & 88.0\% \\
% \bottomrule
% \end{tabular}
% \caption{Calibration of LLM judges against human annotations. Standard-query and
% abstention metrics are reported separately to show how the judges trade off
% standard-query coverage and abstention precision.}
% \label{tab:multi-judge-calibration}
% \end{table}
Table~\ref{tab:abstention-judge-correctness} reports the prediction accuracy of each LLM-as-a-Judge on the 602 human-annotated abstention examples only. This abstention-only sensitivity analysis is separate from the all-sample calibration results in Table~\ref{tab:judge-abstention-ability}, which also includes solvable examples. Each judge evaluates the same outputs independently. \texttt{claude-sonnet-4-5} achieves the highest abstention-only overall prediction accuracy at 87.6\%, followed by \texttt{gemini-2.5-flash}. Judges perform most reliably on \emph{Missing Capability} and \emph{Misrouting} cases, where the decision often depends on whether the requested function or target worker is supported. In contrast, \emph{Ambiguous Query} and \emph{Insufficient Input} remain harder and show larger variation across judges. The confusion matrices in Figure~\ref{fig:abstention-judge-confusion-matrices} help explain this pattern: ambiguous queries are often confused with insufficient-input cases, since both involve underspecified user requests. Insufficient-input cases are also frequently confused with missing-capability or solvable cases, suggesting that judges differ in whether they interpret missing execution details as a request for unsupported functionality, an answerable query, or a query that requires additional user input.

\begin{table}[H]
\centering
\setlength{\tabcolsep}{4pt}
\renewcommand{\arraystretch}{1.12}
\begin{tabular}{lccccc}
\toprule
\textbf{LLM-as-a-Judge} &
\shortstack{\textbf{Overall}\\\textbf{Accuracy}} &
\shortstack{\textbf{Ambiguous}\\\textbf{Query}} &
\shortstack{\textbf{Insufficient}\\\textbf{Input}} &
\shortstack{\textbf{Missing}\\\textbf{Capability}} &
\textbf{Misrouting} \\
\midrule
\texttt{gpt-5} & 79.6\% & 45.5\% & 58.3\% & 87.6\% & 86.8\% \\
\texttt{o3} & 79.6\% & 33.3\% & 66.0\% & 85.6\% & 83.1\% \\
\texttt{claude-sonnet-4-5} & \best{87.6\%} & \best{83.3\%} & \best{76.7\%} & 90.8\% & \best{90.9\%} \\
\texttt{gemini-2.5-flash} & 84.7\% & 41.7\% & 70.9\% & \best{91.5\%} & 85.7\% \\
\bottomrule
\end{tabular}
\caption{Abstention-only judge sensitivity analysis. Each judge evaluates the same 602 human-annotated abstention examples. Category columns report per-category prediction accuracy under the final taxonomy.}
\label{tab:abstention-judge-correctness}
\end{table}

% \section{Abstention Judge Confusion Matrices}
% \label{app:abstention-judge-confusion-matrices}

% Figure~\ref{fig:abstention-judge-confusion-matrices} shows the per-judge confusion
% matrices for final-model abstention predictions. Rows correspond to the
% model-predicted abstention category, and columns correspond to the judge verdict.

\end{document}